\documentclass[letterpaper,twocolumn,showpacs,superscriptaddress,floatfix]{revtex4}
\usepackage[latin1]{inputenc}
\usepackage{bm}
\usepackage{color, amssymb, graphicx, amsfonts,epstopdf }
\usepackage{multirow,amssymb,amsbsy,amsmath,epstopdf}
\usepackage{graphicx}
\usepackage{verbatim}
\usepackage{pifont}
\usepackage{amsmath}
\usepackage{color}


\begin{document}

\title{Strict experimental test of macroscopic realism in a light-matter-interfaced system}

\author{Xiao Liu}
\author{Zong-Quan Zhou$\footnote{email:zq\_zhou@ustc.edu.cn}$}
\author{Yong-Jian Han}
\author{Zong-Feng Li}
\author{Jun Hu}
\author{Tian-Shu Yang}
\author{Pei-Yun Li}
\author{Chao Liu}
\author{Xue Li}
\author{Yu Ma}
\author{Peng-Jun Liang}
\author{Chuan-Feng Li$\footnote{email:cfli@ustc.edu.cn}$}
\author{Guang-Can Guo}
\affiliation{CAS Key Laboratory of Quantum Information, University of Science and Technology of China, Hefei 230026, China}
\affiliation{CAS Center For Excellence in Quantum Information and Quantum Physics, University of Science and Technology of China, Hefei 230026, China}
\date{\today}


\begin{abstract}
Macroscopic realism is a classical worldview that a macroscopic system is always determinately in one of the two or more macroscopically distinguishable states available to it, and so is never in a superposition of these states. The question of whether there is a fundamental limitation on the possibility to observe quantum phenomena at the macroscopic scale remains unclear. Here we implement a strict and simple protocol to test macroscopic realism in a light-matter interfaced system. We create a micro-macro entanglement with two macroscopically distinguishable solid-state components and rule out those theories which would deny coherent superpositions of up to 76 atomic excitations shared by $10^{10}$ ions in two separated solids. These results provide a general method to enhance the size of superposition states of atoms by utilizing quantum memory techniques and to push the envelope of macroscopicity at higher levels.
\end{abstract}

\maketitle

\section{INTRODUCTION}
Quantum mechanics gives us a picture of the world that is so radically counterintuitive that it has changed our perspective on reality itself \cite{EPR,schrodingercat}. Admittedly, when describing the microscopic world, it has been tested in various systems with remarkably excellent agreement between theory and experiments. Nonetheless, it is rather difficult to reconcile the behavior of quantum particles and our intuitive experience when dealing with macroscopic objects, which should occupy definite states at all times and independently of the observers. The question of whether quantum behavior is restricted for large numbers of particles at the macroscopic level by some unknown nonquantum mechanism or contains some limitation that we do not yet understand is fundamentally unresolved \cite{Leggett2002}. Realizing true macroscopic quantum superpositions would constitute one step towards an answer, providing evidence against macroscopic realism (macrorealism).

In their seminal paper \cite{LG1985}, Leggett and Garg (LG) codified our intuition about the macroscopic world into two principles, macroscopic realism \textit{per se} (a macroscopic system must at any time be in a definite one of its macroscopically distinct states) and the possibility to perform noninvasive measurements (measurements that do not influence the actual state or the subsequent system dynamics of the system). Based on these assumptions, they derived a class of inequalities which are used to test for the quantum behavior of a system, called Leggett-Garg inequalities (LGIs). The experimental tests of LGIs have been performed in a wide range of different physical systems spanning from superconducting transmon systems \cite{superconducting1,superconducting2,superconducting3} to photons \cite{photons1,photons2,photons3,photons4}, electron or nuclear spins \cite{espins1,espins2,espins3,nspins1,nspins2,nspins3}, cesium atoms \cite{Cs}, Nd$^{3+}$:YVO$_{4}$ crystals \cite{Nd}, and neutrino oscillations \cite{Neutrino1,Neutrino2}. A review of most of these experiments may be found in Ref. \cite{LGI2014}.

\begin{figure*}[ht]
\centering
\includegraphics[width= 6.0in]{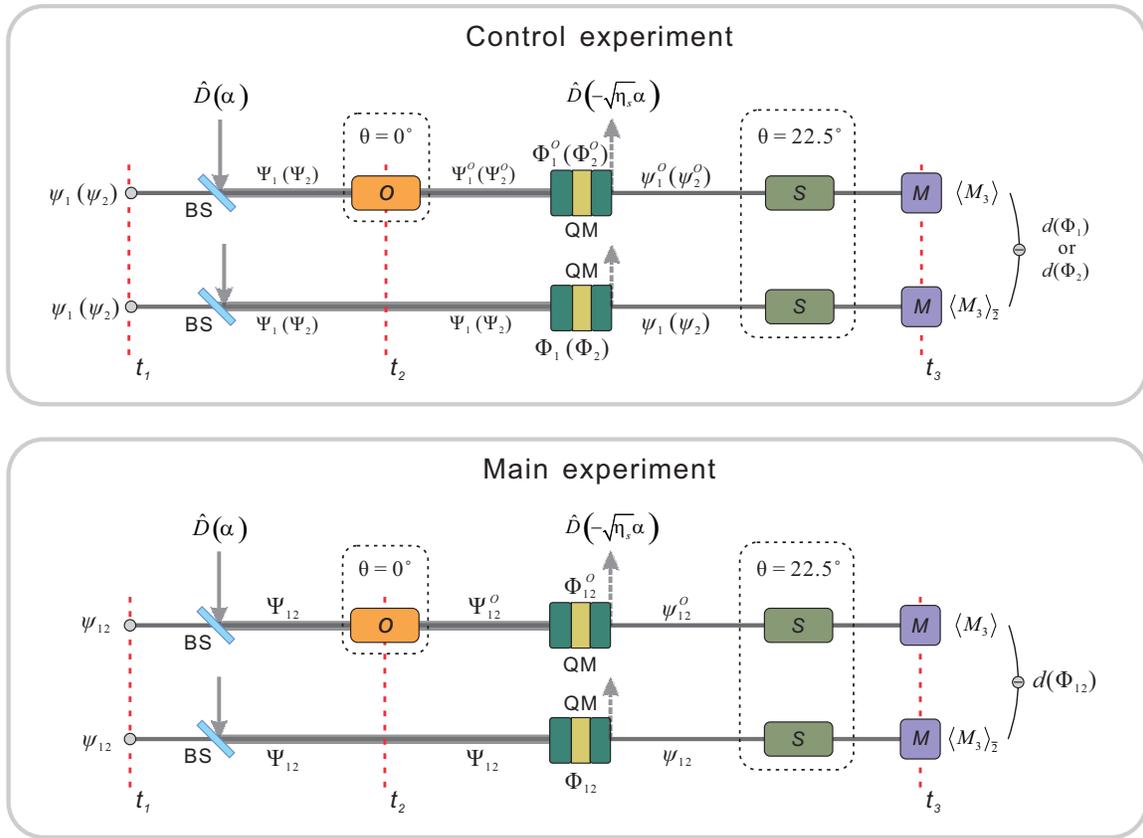}
\caption{Conceptual scheme for the test of macrorealism with a light-matter-interfaced system. Two sets of analogous experiments are performed, one pair of control experiments and one pair of main experiments, depending on the polarization states of single photons prepared at time $t_{1}$. A beam splitter (BS) combining the single photons with coherent light is used to realize displacement operation $\hat{D}(\alpha)$ and a quantum memory (QM) with storage efficiency $\eta_{s}$ is used to map the photon states to atomic states and to create back-displacement operation $\hat{D}(-\sqrt{\eta_{s}}\alpha)$. Here $O$ and $S$ are performed by half waveplates orientated at different angles $\theta$ to realize blind measurement and shuffling operation. The expectation value of a final measurement ($M$) at $t_{3}$ is recorded to show the difference between the presence and absence of $O$ at $t_{2}$. Photon states and atomic states at different times are marked with $\psi$ representing single-photon states, $\Psi$ representing displaced states, and $\Phi$ representing atomic states stored in QM. The subscripts 1 and 2 or 12 stand for two orthogonal classical states or their superpositions starting from the preparation at $t_{1}$.}
\label{NDC}
\end{figure*}

However, there are a number of pitfalls that make the experimental study of LGIs not quite straightforward. For example, LGI tests suffer from the clumsiness loophole that there is the ever-present possibility of a clumsy measurement procedure that gives rise to a violation, rather than any inherent quantum effect. Some clever measurement schemes have been adopted to address this issue, such as ideal negative measurement \cite{Cs,espins3,nspins3} and weak measurements \cite{superconducting1,superconducting2,photons1} or the use of an additional stationarity assumption \cite{STATIONARY,Nd,espins1}. Nevertheless, in principle, no measurements can be treated as truly noninvasive without \textit{a priori} knowledge. More importantly, due to the demanding nature of LGI tests, most of the works have mainly focused on superconducting circuits or small systems such as single atoms or photons, except for a few efforts trying to push the tests to larger objects \cite{LGnew1,LGnew2,LGnew3,LGnew4}. We previously reported a violation of a LGI in a light-matter system \cite{Nd}, of which the millimeter-sized crystals were macroscopic in size and involved a large number of atoms in the delocalized excitation, but the interfering states only differ by the absorption of a single quantum and thus the considered states are of low disconnectivity ($\mathcal{D}=1$), a measure for macroscopicity introduced by Leggett in Ref. \cite{disconnectivity}.

Here we employ a strict test of macrorealism in a light-matter-interfaced system with micro-macro entanglement states. Our implementation, inspired by the proposal of Ref. \cite{micromacro1} and experiments \cite{micromacro2,micromacro3,micromacro4}, follows the spirit of the famous Schr\"{o}dinger cat gedanken experiment \cite{schrodingercat}, which involves a (macroscopic) cat whose quantum state becomes entangled with that of a (microscopic) decaying nucleus. We start with a single-photon micro-micro entanglement, which is subsequently displaced to the micro-macro entanglement using local displacement operation in optical phase space. The displaced photon is then mapped to an atomic ensemble, creating the light-matter micro-macro entangled state. To test macrorealism in such a complex system, we use an experimentally simplified and theoretical strict protocol originated from LG's approach \cite{n1,NIST,espins2,n4,NDC,invasiveness} which focused on determining and accounting for measurement clumsiness instead of directly proving that a measurement is non-invasive.

\section{CONCEPTUAL SCHEME}
Using the same assumptions as LGIs, one can reach a simpler constraint
\begin{equation}\label{ndc}
 d =\langle M_{3}\rangle_{\bar{2}}- \langle M_{3}\rangle = 0,
\end{equation}
called the nondisturbance condition (NDC) \cite{NDC}, which has also been described as a quantum witness \cite{n1,n8} or no signaling in time condition \cite{NIST}. Here $\langle M_{3}\rangle$ and $\langle M_{3}\rangle_{\bar{2}}$ represent measuring the average value of a dichotomic observable $M_{3}$ at time $t_{3}$ with or without a measurement denoted by $O$ at time $t_{2}$. It has been suggested that $O$ can be described as a generalized operation whose properties are to be obtained through experimental setup \cite{NDC}. In other words, we can treat $O$ as a measurement whose result is not inspected \cite{blind}. Compared to original LGI tests, which require us to perform measurements at three different times with measurements of two-point correlations, the NDC is more noise tolerant and can be violated for a much wider parameter regime \cite{NDC}. It greatly reduces the difficulties in testing macrorealism in large and complex systems.

The conceptual scheme is shown in Fig. \ref{NDC}. Unlike the LGI tests implemented previously, in which \textit{a priori} arguments that the invasiveness of the measurement is zero have been employed, we adopt another approach to determine the invasiveness of the measurement \cite{invasiveness,NDC}. The whole NDC test requires two sets of analogous experiments, i.e. one pair of control experiments used to determine the worst case disturbance when classical states are prepared and one pair of main experiments to measure the disturbance not explainable merely by appealing to the clumsiness revealed in the control experiments. A photon from a single-photon source is first prepared in horizontal polarization ($|H\rangle$), vertical polarization, ($|V\rangle$), or their superposition. In the main experiments, we choose $\frac{1}{\sqrt{2}}(|H\rangle+|V\rangle)$ for maximum violation of macrorealism (see the Appendix for details). The state of the photon can be written as a simple form of entanglement between polarization modes
\begin{equation}\label{micromicro}
  |\psi_{12}\rangle = \frac{1}{\sqrt{2}}(|1\rangle_{H}|0\rangle_{V}+|0\rangle_{H}|1\rangle_{V}),
\end{equation}
known as single-photon entanglement \cite{spe,micromacro1}, where $|1\rangle_{H}|0\rangle_{V}=|\psi_{1}\rangle $ and $ |0\rangle_{H}|1\rangle_{V}=|\psi_{2}\rangle$ correspond to $|H\rangle$ and $|V\rangle$, respectively.

\begin{figure*}[ht]
\centering
\includegraphics[width= 6.5in]{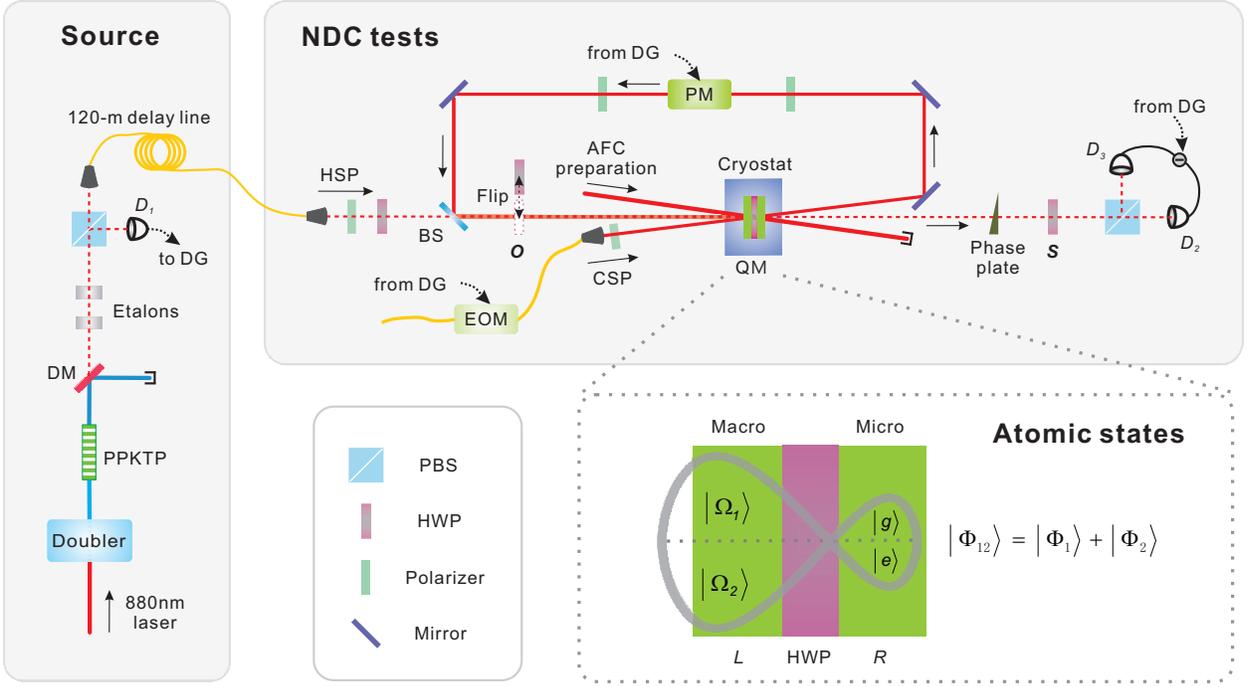}
\caption{Experimental setup. The source is a PPKTP crystal is pumped by a 440-nm laser which is frequency doubled from a continuous-wave Ti:sapphire laser at 880 nm. Pairs of photons are generated by a type-II spontaneous parametric down-conversion process and separated by a polarizing beam splitter (PBS) after spectrally filtered to a linewidth of approximately 700 MHz by two etalons. The pump light is removed by the dichroic mirror (DM). The idler photons are detected by a single-photon detector (SPD) denoted by $D_{1}$. One click of $D_{1}$ heralds a signal photon and triggers a delay generator (DG), which further triggers the following experiment sequence such as the generation of coherent state pulse (CSP) and the coincidence with heralded signal photon (HSP). The CSP should be generated before the HSP reaches the BS and therefore the HSP is delayed by a 120-m fiber line. For the NDC tests the polarization of the HSP is prepared through a HWP. The CSP is generated by an electro-optical intensity modulator (EOM) and sent into the QM in a different spatial mode than the signal photon. The transmission part of the CSP is synchronized with the HSP on a BS that has a 99.3\% transmittance, corresponding to a displacement operation $\hat{D}(\alpha)$. The resulting micro-macro state is stored inside the QM after passing through a $0^{\circ}$ HWP fixed on a flip mount, which is used to decide whether to perform operation $O$ or not and released after a storage time of 50 ns. We apply a $\pi$ phase shift on the storage part of the CSP via a electro-optical phase modulator (PM), which is then applied on the state retrieved from the QM as a back-displacement operation. The state is analyzed with a HWP at $22.5^{\circ}$ that acts as an $S$ operation followed by a PBS and two SPDs ($D_{2}$ and $D_{3}$) that act as a measurement on a polarization basis. For the atomic states, after being mapped in the QM, the displaced photon state can been seen as a micro-macro entanglement of excitation states between two separated crystals with the macro component in the $L$ crystal and the micro component in the $R$ crystal. It can also be seen as a superposition of two macroscopic distinguishable states $|\Phi_{1}\rangle=|\Omega_{1}\rangle_{L}|g\rangle_{R}$ and $|\Phi_{2}\rangle=|\Omega_{2}\rangle_{L}|e\rangle_{R}$.}
\label{setup}
\end{figure*}


To displace one of the polarization modes of the single photon, the photon is superimposed on a horizontally polarized bright coherent state on a highly asymmetric beam splitter \cite{BS1,BS2,micromacro1,micromacro2,micromacro3,micromacro4}. This corresponds to a unitary displacement operation $\hat{D}(\alpha)$ on the horizontal mode of the photon. The average number of photons contained in the displacement pulse is given by $|\alpha|^{2}$. After displacement, the state is written as
\begin{equation}\label{micromacro}
   |\Psi_{12}\rangle = \frac{1}{\sqrt{2}}(\hat{D}(\alpha)|1\rangle_{H}|0\rangle_{V}+|\alpha\rangle_{H}|1\rangle_{V}),
\end{equation}
where $\hat{D}(\alpha)|1\rangle_{H}$ represents a displaced single-photon state which can be characterized by a photon-number distribution with a mean photon number $|\alpha|^{2}+1$ and a variance $3|\alpha|^{2}$; $|\alpha\rangle_{H}$ is a coherent state, which follows a Poissonian photon-number distribution with mean photon number $|\alpha|^{2}$ equal to the variance. Although these two terms differ by only one photon on average, the distance between their photon-number distributions is $|\alpha|$. Thus, by checking whether the photon number falls in the range $|\alpha|^{2}\pm|\alpha|$ using a classical coarse-grained detector, one can distinguish their photon-number distributions with a probability approaching 74\% for large enough $\alpha$ \cite{micromacro2}. More intriguingly, if the state of Eq. (\ref{micromacro}) is seen in another basis so that it reads $\hat{D}(\alpha)(|0\rangle_{H}+|1\rangle_{H})(|0\rangle_{V}+|1\rangle_{V})-\hat{D}(\alpha)(|0\rangle_{H}-|1\rangle_{H})(|0\rangle_{V}-|1\rangle_{V})$, where the normalization is omitted, the probability to distinguish the states $\hat{D}(\alpha)(|0\rangle_{H}+|1\rangle_{H})$ and $\hat{D}(\alpha)(|0\rangle_{H}-|1\rangle_{H})$ in a single shot with a classical measurement can reach up to 90\% \cite{Size}. For this reason, the state (\ref{micromacro}) can be considered as a micro-macro entangled state with the horizontal mode of the signal photon playing the role of the `macro' component and the vertical mode of the micro component.

We use a half waveplate (HWP) oriented at $0^{\circ}$ to perform operation $O$. After $O$, the state of Eq. (\ref{micromacro}) becomes $ |\Psi_{12}^{O}\rangle =\frac{1}{\sqrt{2}}(|\Psi_{1}\rangle-|\Psi_{2}\rangle)$. In principle, $O$ does not affect $|\Psi_{1}\rangle = \hat{D}(\alpha)|1\rangle_{H}|0\rangle_{V}$ and $|\Psi_{2}\rangle = |\alpha\rangle_{H}|1\rangle_{V}$, which can be seen as classical states instead of quantum superposition states.

Then we use the atomic frequency comb (AFC) storage protocol \cite{AFC} to coherently map the state of the displaced photon to the collective state of atoms in an optical quantum memory (QM). This is achieved by the interaction of light with an ensemble of atomic absorbers with an inhomogeneously broadened absorption line that has been tailored into a series of equally spaced absorption peaks. The absorption of photons leads to collective excitations shared by $10^{10}$ atoms. The memory hardware is composed of two pieces of Nd$^{3+}$:YVO$_{4}$ crystal sandwiching a $45^{\circ}$ HWP. The $H$- and $V$-polarization components of light with a wavelength of approximately 880 nm can be independently processed by the first (denote as $L$) and second (denote as $R$) crystals, respectively. This configuration was previously employed for reliable storage of photonic qubits \cite{QM}.

After absorption by the QM, the state, assumed with no operation $O$ applied, of Eq. (\ref{micromacro}) can be written as an atomic state
\begin{equation}\label{atomicstate}
  |\Phi_{12}\rangle=|\Omega_{1}\rangle_{L}|g\rangle_{R}+|\Omega_{2}\rangle_{L}|e\rangle_{R},
\end{equation}
where $|g\rangle_{R}$ and $|e\rangle_{R}$ denote states for the $R$ crystal with and without one-photon excitation, while $|\Omega_{1}\rangle_{L}$ and $|\Omega_{2}\rangle_{L}$ represent the atomic excitation states of $\hat{D}(\alpha)|1\rangle$ and $|\alpha\rangle$ in the $L$ crystal. We define the dichotomic observable $M_{A}=|\Phi_{1}\rangle\langle\Phi_{1}|+|\Phi_{2}\rangle\langle\Phi_{2}|$, where the atomic state $|\Phi_{1}\rangle=|\Omega_{1}\rangle_{L}|g\rangle_{R}$ defines the basis state with eigenvalue equal to $+1$ and the orthogonal state $|\Phi_{2}\rangle=|\Omega_{2}\rangle_{L}|e\rangle_{R}$ corresponds to the basis state with eigenvalue equal to $-1$. To probe the atomic state, the atomic excitations are converted back to the optical mode and at the same time displaced back to the single-photon entanglement state for further measurement. A similar method has been employed in preparing micro-macro entanglement states and inferring quantum correlations by entanglement witness \cite{micromacro4}. It is worth noting that, in principle, a single macroscopic quantum state is necessary to test macrorealism instead of the entanglement state. Therefore, in our system, what we really need are two macroscopic branches of the superposition. The reason why we use the micro-macro entanglement is mainly because it is easier to create and measure by displacement operation, which allows us to test macrorealism in an uncomplicated way, though indirectly. Specifically, the measurement of atomic states $M_{A}$ is indirectly measured by detecting back-displaced photon states at time $t_{3}$ which is in the basis of $|\psi_{1}\rangle=|1\rangle_{H}|0\rangle_{V}$ and $|\psi_{2}\rangle=|0\rangle_{H}|1\rangle_{V}$, which we define to be $M_{3}$.

\section{EXPERIMENTAL SETUP AND RESULTS}
The experimental setup is illustrated in Fig. \ref{setup}. The detection of one idler photon heralds the creation of a signal photon, which is sent to the light-matter system through an optical fiber for NDC tests.

The 5-ppm doped Nd$^{3+}$:YVO$_{4}$ crystals are placed in a cryostat at a temperature of 1.7 K and with a magnetic field of 0.3 T. Light incident on the QM is either absorbed with probability $\eta_{abs}=92.2(1)\%$ or transmitted with probability $\eta_{t}=1-\eta_{abs}=7.8(1)\%$. The absorbed light is re-emitted after a predetermined time $\tau_{s} = 50$ ns from QM with probability $\eta_{s}=18.3(2)\%$. The state retrieved from the QM is immediately displaced back to the micro-micro entangled state with $\hat{D}(-\sqrt{\eta_{s}}\alpha)$, where the amplitude is reduced by $\sqrt{\eta_{s}}$ to match the limited storage efficiency $\eta_{s}$ of the QM \cite{micromacro4}. To explain how the QM can be used to realize the back-displacement operations, let us return to the generation of the coherent state pulse (CSP), which is triggered whenever an idler photon is detected. It corresponds to the displacement $\hat{D}(\sqrt{1/\eta_{t}}\alpha)$ applied on vacuum. The CSP is sent through the QM before the signal photon arrives at the beam splitter (BS). This creates two pulses, a directly transmitted pulse and a stored pulse delayed by the storage time $\tau_{s} = 50$ ns of the QM, which corresponds to the displacements $\hat{D}(\alpha)$ and $\hat{D}(\sqrt{\eta_{s}/\eta_{t}}\alpha)$, respectively. The transmitted pulse is superimposed on the heralded signal photon (HSP) on the BS, displacing the micro-micro entangled state $\psi$ to the micro-macro entangled state $\Psi$.  We apply a phase of $\pi$ on the storage part of the CSP via a free-space phase modulator (PM) [the corresponding displacement operation becomes $\hat{D}(-\sqrt{\eta_{s}/\eta_{t}}\alpha)$]. The back-displacement happens when the storage part of CSP transmits QM again, overlapped precisely in time with the retrieved signal of $\Psi$ \cite{micromacro4}. To quantify the quality of the back-displacement, we measured the visibility between the interference of two displacements with or without PM working when the signal photon is blocked. A visibility of 99.8\% is obtained.

\begin{figure}[tb]
\centering
\includegraphics[width= 0.95 \columnwidth]{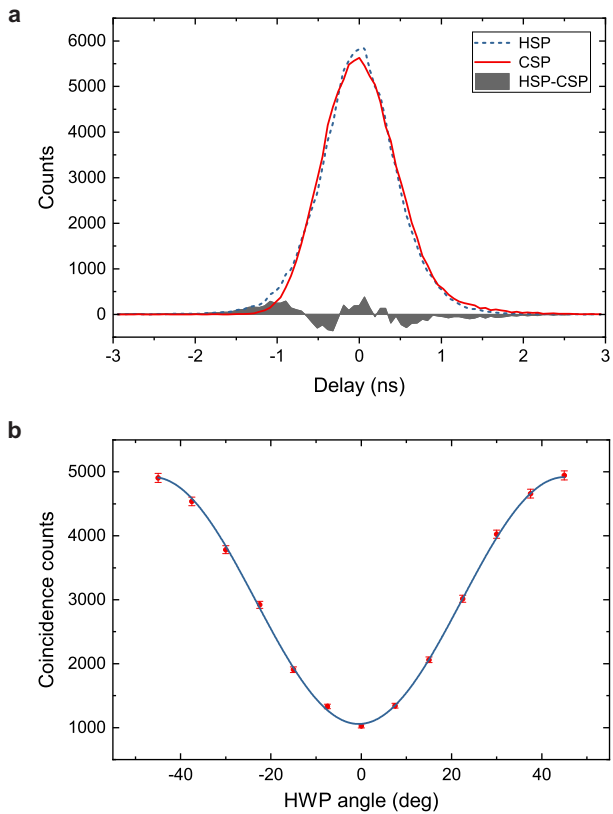}
\caption{(a) Temporal profiles of the coherent state pulse and heralded signal photon. The dashed blue line is the temporal mode of the CSP shaped from the continuous-wave laser using the EOM and the solid red line is the temporal profile of the HSP. The gray shaded area represents the difference between two profiles corresponding to the CSP and HSP. A coincidence window of 3 ns (dashed rectangle) was used for further analysis. The likeness between these two pulses is calculated to be 99.6\%. (b) HOM dip measurement between the HSP and CSP. By rotating a HWP that is used to control polarization of the HSP, we can extract the visibility of the HOM dip. The red dots are the coincidence rate as a function of the angle of the HWP. The solid blue line is the theoretical fitting of experimental data. The error bars are $\pm1$ standard deviation estimated from Monte Carlo simulations based on the Poisson statistics of photon counts.}
\label{HOM}
\end{figure}

We tune the pulse width and arriving time of the CSP superimposed on the HSP carefully with an arbitrary waveform generator. However, the overlapping is still not perfect [see Fig. \ref{HOM}(a)]. Therefore, we must evaluate to what extent their modes are indistinguishable. This was done by utilizing the Hong-Ou-Mandel (HOM) type of interference. The HSP and CPS are combined on a 54:46 BS and synchronized in time. We measure the coincidence rate at two output ports of the BS as we change the polarization of the HSP through a HWP. A two-photon interference can be observed when the polarization of the HSP is parallel to the CSP and as we change the polarization from parallel to perpendicular, the photons from two sources become more and more distinguishable so that the interference gradually disappears. A HOM dip is shown in Fig. \ref{HOM}(b) for the mean photon number of the CSP $|\alpha|^{2}=0.01$. The measured visibility is $V_{m} = 78.5(8)\%$. We also calculate the expected visibility $V_{e} = 92.1\%$, taking into account the heralding efficiency and the photon pair creation probability of the HSP (see the Appendix for the detailed calculation). The measured ratio $R = V_{m}/V_{e} = 85.2\%$ is used to correct the size of the displacements that are presented in this work.

\begin{figure}[ht]
\centering
\includegraphics[width= 0.95 \columnwidth]{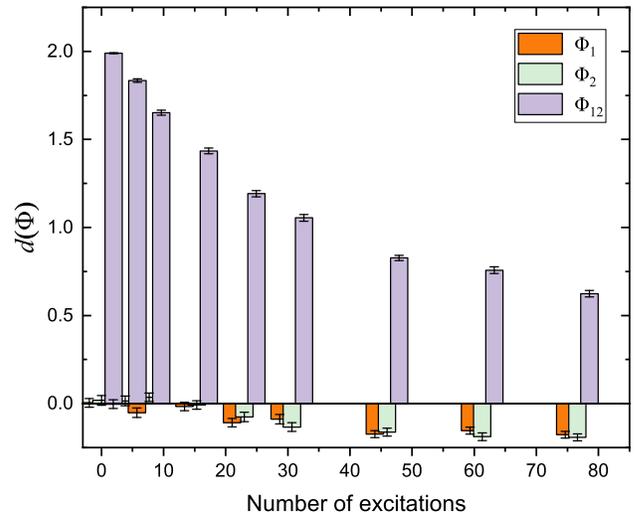}
\caption{ Experimental violation of the nondisturbance condition. The disturbance parameter $d(\Phi)$ is plotted as a function of the number of excitations stored in QM. Three different colors of histograms represent the disturbance parameter of $d(\Phi_{1})$ and $d(\Phi_{2})$ revealed in the control experiments and $d(\Phi_{12})$ revealed in the main experiments. The error bars are $\pm1$ standard deviation estimated from Monte Carlo simulations based on the Poisson statistics of photon counts.}
\label{MAIN}
\end{figure}

Before we present the experimental results of testing macrorealism by violation of the NDC with the above-mentioned micro-macro entangled states, we need to define a precise and operational notion of macrorealism that is tested in our experiment. Using conditional probabilities, the disturbance parameter can be defined as \cite{NDC}
\begin{equation}\label{drou}
\begin{aligned}
  d(\Phi)= & [P(M_{3}=+1|\Phi)-P(M_{3}=-1|\Phi)] \\
  &-[P(M_{3}=+1|\Phi,O)-P(M_{3}=-1|\Phi,O)],
  \end{aligned}
\end{equation}
as a measure of how much disturbance is introduced to $M_{3}$ by applying $O$ at $t_{2}$ (compared with doing nothing) when the state mapped in QM is described by state $\Phi$. In a pair of control experiments, we determine $d(\Phi_{1})$ and $d(\Phi_{2})$, where $\Phi_{1}$ and $\Phi_{2}$ are the states for which the measurement reveals classical disturbance. Once the control experiments are completed, the main experiments may begin to determine $d(\Phi_{12})$. According to the NDC, the fact that $d(\Phi_{1}) = d(\Phi_{2})=0$ but $d(\Phi_{12})\neq0$ could be thought of as a violation of the macrorealist view.

The disturbance parameter measured at different average number of atomic excitations inside the QM is shown in Fig. \ref{MAIN}. We find violations $|d(\Phi_{12})|-|d(\Phi_{1})|=0.44692$ and $|d(\Phi_{12})|-|d(\Phi_{2})|=0.43221$ , which are both 11 standard deviations away from zero for up to 76 excitations. The classical disturbances $d(\Phi_{1})=-0.17692$ and $d(\Phi_{2})=-0.19166$ deviate from $0$. This is mainly due to the noise caused by the bright CSP. Our results still show a violation of the NDC by more than two standard deviations even if we treat $|d(\Phi_{1})|$ and $|d(\Phi_{2})|$ as total deviations of the measurement.

In our experiment, the maximum number of excitations stored in QM is 76, which corresponds to $|\alpha|^{2}=83$ before QM. Although the two macroscopic components are not totally distinguishable, we can still evaluate the size of the macroscopic state from the coarse-grained measure described in Sec. \uppercase\expandafter{\romannumeral2}. Based on this method, the effective size of micro-macro entangled photon state before QM can be determined by first quantifying the maximal amount of noise that still allows one to distinguish $\hat{D}(\alpha)(|0\rangle_{H}+|1\rangle_{H})$ and $\hat{D}(\alpha)(|0\rangle_{H}-|1\rangle_{H})$ with a fixed probability $P_{g}$ and then comparing this to an archetypical state involving the superposition of vacuum state $|0\rangle$ and $N$-photon Fock state $|N\rangle$, which we calibrate to be of size $N$ \cite{Size}. For $P_{g}=2/3$, the maximum micro-macro entangled photon state obtained in our experiment before the QM is analogous to the state $1/\sqrt{2}(|\uparrow\rangle|0\rangle+|\downarrow\rangle|N\rangle)$ with $N=5.14$, where $|\uparrow\rangle$ and $|\downarrow\rangle$ represent microscopic orthogonal states. According to its definition, the disconnectivity should measure the number of particles that behave differently in the two branches of the superposition. Hence, taking account of the absorption probability of the QM, the maximum disconnectivity of the atomic states achieved in our system can be estimated to be $\mathcal{D}= \eta_{abs} N \approx5$ \cite{disconnectivity}. For comparison, $\mathcal{D}$ realized in the previous demonstration of light-matter entanglement is approximately 3 \cite{micromacro4}.  A violation of the NDC reported in the micrometer-sized superconducting system has realized a $\mathcal{D}$ of 8 \cite{NDC}. The current system has the advantages of increased size and complexity, as well as long-lived coherence for macroscopic superposition states.

\section{CONCLUSION}
In this work we have reported a strict experimental test of macroscopic realism in a light-matter interfaced system with up to 76 atomic excitations. Our results provide evidence for the superposition of states of nontrivial macroscopic distinctness that violates the macrorealist bound with a high degree of statistical significance, though in an indirect way by detecting the states of single photons. These atomic states can in principle be directly distinguished using a readout technique that has an intrinsically limited microscopic resolution, as it was shown experimentally in Ref. \cite{direct}. Two main reasons attribute the limitation of the size that could be tested in the experiment. First, the imperfectness of the AFC used for storage of the bright CSP brings undesired noise in the detection window. Second, the counting rate of single-photon detectors (SPDs) for signal photons reaches the dead time limit when the size of the displacement is too large. We note that these problems can be addressed with various developing techniques, for example, the creation of a near-perfect AFC \cite{perfectAFC1,perfectAFC2} and the use of a SPD with a high repetition rate \cite{SPD}. Another intriguing possibility is to combine our approach with homodyne detection \cite{micromacro3,homo2}, which should greatly expand the size of quantum superpositions in matter and may enable the tests at higher levels of macroscopicity.

\section*{ACKNOWLEDGMENTS}
We are most grateful to C. Emary and G. Knee for their helpful comments. This work was supported by the National Key R\&D Program of China (Grant No. 2017YFA0304100), the National Natural Science Foundation of China (Grant No. 61327901, No. 11774331, No. 11774335, No. 11504362, No. 11654002, and No. 11821404), Anhui Initiative in Quantum Information Technologies (Grant No. AHY020100), Key Research Program of Frontier Sciences, CAS (Grant No. QYZDY-SSW-SLH003), the Fundamental Research Funds for the Central Universities (Grant No. WK2470000023 and No. WK2470000026).

\section*{APPENDIX}
\subsection{Details on the experimental setup}
The pump light used for the preparation of the AFC is generated through an acousto-optic modulator in double-pass configurations, in which the frequency is swept over 100 MHz in a 500-$\mu$s cycle and each frequency point has been assigned a specific amplitude to give a comb structure. The bandwidth of the AFC is further extended to over 700 MHz using a fiber-pigtailed electro-optic phase modulator. To protect the SPDs (including $D_{1}$ and $D_{2}$) during the preparation procedure, two phase-locked mechanical choppers are placed in the pumping optical path and before the SPDs, respectively. The pump light, the HSP, and the CSP are overlapped at the sample with a noncollinear configuration. The HSP and CSP focus to a diameter of 100 $\mu$m, while the pump light is collimated to produce a much larger diameter on the sample. The signal from $D_{1}$ and $D_{2}$ is sent to the time-correlated single-photon counting system (PicoQuant, HydraHarp 400). The detection efficiency is approximately 0.256, taking into account the detection efficiency of the SPDs ($\sim 0.4$) and transmittance from the sample to the detectors ($\sim 0.64$). The timing sequence used for the storage is controlled by two arbitrary function generators (Tektronix, AFG3252). The AFC preparation takes 12.5 ms. To avoid the fluorescence noise caused by the classical pump light, the measurement cycle starts 1.5 ms after the preparation completes. Photon pulses are stored in the sample in the 10-ms measurement cycle. The complete preparation and measurement cycles are repeated at a frequency of 40 Hz.

The CSP is generated by a fiber-pigtailed electro-optical intensity modulator that carves a pulse out of a continuous-wave laser. The extinction ratio of the modulator is approximately $8000:1$. We use an external DC bias control circuit to maintain the stability of the extinction ratio. The pulse shape of the CSP is controlled by a 8-GS/s arbitrary waveform generator (Tektronix, AWG7082C) and carefully tuned to obtain high likeness with the HSP.

\subsection{Maximum violation of macrorealism}
In the main experiments, we choose the angle of the HWP that is used to prepare the input states to be $22.5^{\circ}$ for maximum violation of macrorealism. To explain this, we first assume the angle of the HWP to be $\theta$. The polarization state of a signal photon becomes $\cos(2\theta)|H\rangle+\sin(2\theta)|V\rangle$, which can be rewritten as a single photon entanglement state
\begin{equation}\label{micromicro2}
  |\psi_{12}\rangle = \cos(2\theta)|1\rangle_{H}|0\rangle_{V}+\sin(2\theta)|0\rangle_{H}|1\rangle_{V}.
\end{equation}

We now focus on the photon state retrieved from QM. If there is no operation $O$ imposed on the displaced state, the state should be still $|\psi_{12}\rangle$; if there is $O$, the retrieved state becomes
\begin{equation}\label{micromicro2}
  |\psi_{12}^{O}\rangle = \cos(2\theta)|1\rangle_{H}|0\rangle_{V}-\sin(2\theta)|0\rangle_{H}|1\rangle_{V}.
\end{equation}

The angle of the HWP that serves as shuffling operation $S$ is set to $22.5^{\circ}$ for realization of the NDC with classical atomic states $|\Phi_{1}\rangle$ and $|\Phi_{2}\rangle$. After $S$, states $|\psi_{12}\rangle$ and $|\psi_{12}^{O}\rangle$ are changed to
\begin{equation}\label{micromicro2}
  |\psi_{12}^{O}\rangle_{S_{2}} = \cos(45^{\circ}+2\theta)|1\rangle_{H}|0\rangle_{V}+\sin(45^{\circ}+2\theta)|0\rangle_{H}|1\rangle_{V}
\end{equation}
and
\begin{equation}\label{micromicro2}
  |\psi_{12}\rangle_{S_{2}} = \cos(45^{\circ}-2\theta)|1\rangle_{H}|0\rangle_{V}+\sin(45^{\circ}-2\theta)|0\rangle_{H}|1\rangle_{V}.
\end{equation}

Measurement of the observable $M_{A}$ of the atomic state that is in the basis of $|\Phi_{1}\rangle$ and $|\Phi_{2}\rangle$ is equivalent to the measurement of the observable of the corresponding retrieved single-photon entanglement state that is in the basis of $|\psi_{1}\rangle=|1\rangle_{H}|0\rangle_{V}$ and $|\psi_{2}\rangle=|0\rangle_{H}|1\rangle_{V}$.

According to the NDC, the disturbance parameter $d(\Phi)= [P(M_{3}=+1|\Phi)-P(M_{3}=-1|\Phi)]-[P(M_{3}=+1|\Phi,O)-P(M_{3}=-1|\Phi,O)]$ should always be $0$ \cite{NDC}. Here we obtain
\begin{equation}\label{drou}
\begin{aligned}
  d(\Phi)=&[|\cos(45^{\circ}-2\theta)|^{2}-|\sin(45^{\circ}-2\theta)|^{2}]-\\
  &[|\cos(45^{\circ}+2\theta)|^{2}-|\sin(45^{\circ}+2\theta)|^{2}]=2\sin(4\theta).
    \end{aligned}
\end{equation}
It is clear that when $\theta=22.5^{\circ}$, theoretically, we get a maximum violation of the NDC of $d(\Phi_{12})=2$.

\subsection{Indistinguishability between the two photon sources}
Due to the possibility of multiphoton creation for both the HSP and CSP, the theoretical visibility of the HOM dip is not unitary, which is given by \cite{micromacro2}
\begin{equation}\label{visibility}
  V_{e}=\frac{P_{1,1}}{P_{2,0}+P_{0,2}+\frac{r^{2}+t^{2}}{2rt}P_{1,1}},
\end{equation}
where $P_{i,j}$ represents the probability to have $i$ and $j$ photons at the input ports of the BS characterized by a transmission $t$ and a reflection $r$. In our experiment, the HSP source is characterized by a heralding efficiency $\eta_{h}=0.065$ and a photon pair creation probability $p = 0.00003$ in a 3-ns detection window. By setting the mean number of photons in the CSP at 0.01 photons per pulse, the expected theoretical visibility is $V_{e} = 92.1\%$ for a 54:46 BS. We obtain an experimental visibility $V_{m} = 78.5(8)\%$. The difference between theoretically calculated visibility and experimentally measured visibility is mainly due to the imperfect overlapping between the HSP and CSP in their temporal modes, which indicates that single-photon entanglement states are not fully displaced by the CSP. Thus, the ratio $R = V_{m}/V_{e}$ is used to correct the size of the displacements that are presented in this work. Namely, if the size of the CSP is $|\alpha|^{2}$, then the size of the displacement is $R|\alpha|^{2}$.


\end{document}